\documentclass{ws-procs11x85}
\newcommand{\abs}[1]{\abstract{#1}}

\usepackage{amsmath}
\usepackage{amssymb}

\usepackage{xspace}
\usepackage{cite}

\newcommand{\LLx}{LL$x$\xspace}
\newcommand{\NLLx}{NLL$x$\xspace}

\newcommand{\ie}{\emph{i.e}.\xspace}
\newcommand{\eg}{\emph{e.g.}\xspace}
\newcommand{\cnf}{\emph{cf.}\ }

\newcommand{\NLLB}{NLL$_\mathrm{B}$\xspace}
\newcommand{\as}{\alpha_s}              % coupling constant
\newcommand{\asb}{\bar{\alpha}_s}       % alpha_s bar
\newcommand{\order}[1]{\mathcal{O}\left(#1\right)}

\newcommand{\nf}{n_f}
\newcommand{\NC}{N_c}
\newcommand{\MSbar}{\overline{\mbox{\scriptsize MS}}}

\begin{document}

% LPTHE-P05-02

\title{Asymptotics and preasymptotics at small $x$%
  \thanks{\lowercase{\uppercase{T}alk presented at the \uppercase{QCD}
      at cosmic energies workshop, \uppercase{E}ttore
      \uppercase{M}ajorana \uppercase{c}entre, 
      \uppercase{E}rice, \uppercase{I}taly, \uppercase{S}eptember 2004.}}}

%Instructions for Producing a Camera-Ready Manuscript 
%using Latex for Publication in Conference 
%Proceedings\footnote{\uppercase{T}his work is supported by etc, etc.}}

\author{G.~P. SALAM}
%\footnote{\uppercase{W}ork partially
%supported by grant 2-4570.5 of the \uppercase{S}wiss 
%\uppercase{N}ational \uppercase{S}cience \uppercase{F}oundation.}}

\address{LPTHE, Universities of Paris VI and VII and CNRS,\\
  75005 Paris, France\\
E-mail: salam@lpthe.jussieu.fr}

%\author{T.~R. SIMON, S. CLARKE and S.~N. GERALD}
%
%\address{World Scientific Publishing Co Ltd, \\ 
%57 Shelton Street, \\
%London WC2H 9HE, England\\
%E-mail: wspc@wspc.ox.uk}  

\maketitle

\abs{This talk discusses the relative impact of running-coupling
  and other higher-order corrections on the small-$x$ gluon-gluon
  splitting function.  Comments are made on similarities with some
  aspects of the Balitsky-Kovchegov equation, which arise because of
  the presence of an effective infrared cutoff in both cases. It is
  emphasised that, at least in the splitting-function case, the
  asymptotic small-$x$ behaviour has little relevance to the
  phenomenologically interesting preasymptotic region. This is
  illustrated with the aid of a convolution of the resummed splitting
  function with a toy gluon distribution.}

%----------------------------------------------------------------------
\section{Introduction}

Detailed introductions to the more theoretical aspects of small-$x$
physics have been given in the contributions to these proceedings by
Ciafaloni \cite{CiafaloniTheseProc} and by Mueller
\cite{MuellerTheseProc}. The former concentrated on our understanding
of the all-orders perturbative structure of the linear problem of
small-$x$ parton multiplication, while the latter discussed the new
phenomena that occur when the gluon density becomes so high that the
small-$x$ growth \emph{saturates}.

In the linear regime there have been extensive studies of the
higher-order corrections. These are essential, insofar as the
leading-logarithmic (\LLx) BFKL equation \cite{BFKL} for small-$x$
growth is clearly inconsistent with data (see for example
\cite{BF95,EHW}). However, the pure next-to-leading logarithmic (\NLLx)
contributions to the evolution, \cite{NLLFL,NLLCC}, are so large that
the problem appears perturbatively unstable.\footnote{Though there are
  certain observables for which specific implementations of the pure
  \NLLx corrections may have reduced instabilities
  \cite{Andersen-SabioVera}.} Techniques have been developed over the
past few years to help understand the origin of the poor perturbative
convergence, in the hope that one may then use that understanding
to help reorganise the perturbative series into a more stable hierarchy.
As discussed in \cite{CiafaloniTheseProc}, methods based on the
combination of collinear and small-$x$ resummations
\cite{Salam1998,CC,CCS1,CCS00,CCSSkernel,ABF2000,ABF2001,ABF2003,ABFcomparison,ABF2004,FRSV}
seem to be particularly successful in this respect.

The situation in the context of saturation studies is less developed.
Firstly, there is no definitive understanding of how to extend linear
\LLx BFKL evolution to the saturation regime. One of the most widely
studied models is the Balitsky-Kovchegov (BK) equation
\cite{Balitsky,Kovchegov}, which can be understood as resumming
pomeron fan diagrams \cite{GLR}, and for which an additional
mean-field approximation is nearly always made (formally valid only
for a thick nucleus, and over a limited energy range). Other more
sophisticated approaches to saturation are currently being
investigated (\eg
\cite{MuellerShoshi,IMM,IancuTriant,LevinLublinsky}), a number of
which aim to account for pomeron loops, first shown to be important in
some early numerical calculations \cite{Salam1995,MuellerSalam} within
the dipole approach \cite{MuellerDipole}. Secondly, even within the
simplest, BK, approach to saturation, studies of higher-order
corrections have been less extensive than for the linear BFKL
equation.

One purpose of this talk is to examine some general lessons that have
been learnt about the effects of higher-order corrections in the case
of linear evolution and to discuss how they might be relevant also in
the BK saturation case.

A second part of this talk will examine briefly the outlook for
phenomenological applications of the higher-order linear BFKL
framework.

%----------------------------------------------------------------------
\section{General aspects of higher-order BFKL corrections}

In order to discuss higher-order corrections in linear and saturating
BFKL, it is important to understand what precisely to compare. A
critical feature of the BK equation in this context is that its
non-linear term provides an effective transverse infrared cutoff on
the evolution, usually known as the saturation scale $Q_s$. This
cutoff scale varies as a function of $y = \ln 1/x$. A very elegant
formalisation of these properties has recently been given in
\cite{MunierPesch}.

\begin{figure}[htbp]
  \centering
  \includegraphics[width=3.6in]{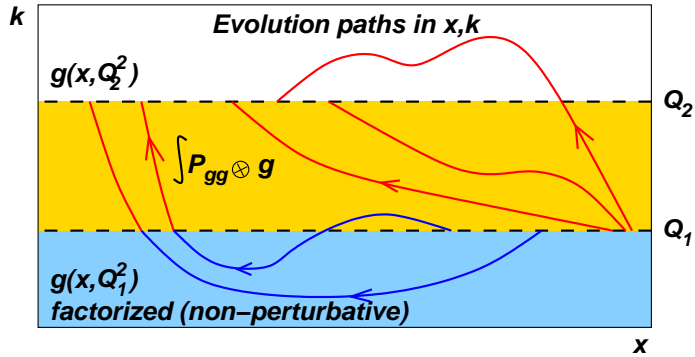}
  \caption{Evolution paths between two transverse scales, $Q_1$ and
    $Q_2$, and their separation into the parton distribution (paths
    that go below $Q_1$, blue if viewed in colour) and the
    perturbative DGLAP evolution (remaining paths, red).  }
  \label{fig:paths}
\end{figure}

Infrared cutoffs have long been investigated in linear BFKL.  They
arise (a) when one imposes them ad-hoc to eliminate the
non-perturbative regime, or (b) implicitly, in the study of the
gluon-gluon splitting function $P_{gg}(z)$, which through
factorisation contains just ultraviolet evolution, while infrared
evolution (that below the factorisation scale) is entirely in the
gluon distribution function, $g(x,Q^2)$, as illustrated in
figure~\ref{fig:paths} (see \eg \cite{THORNE,CCS00,ABF2001}).

In all BFKL-type problems the question of higher-order corrections is
doubly complicated, because in addition to the usual \NLLx corrections
(relative order $\as$ compared to \LLx), the iteration of the kernel
means that it is not possible to identify a unique scale at which to
evaluate the kernel. In problems without cutoffs (and with two probes
at similar hard scales), it turns out that it is nevertheless a
reasonable approximation to use the hard scale of the problem as the
effective scale, at least over a moderately large range of $Y$,
because fluctuations in scale due to BFKL diffusion are, to a first
approximation, symmetric around the hard scale
\cite{KovchegovMuellerRunning,CTM,CCSSHardPom}. In contrast, in each
of the three cutoff-contexts mentioned above, it was discovered,
independently \cite{hr92,CCS1,MuellerTrianta}, that if one uses the
cutoff scale (the only physically unambiguously identifiable scale) as
the renormalisation scale, then there are large negative corrections
to the BFKL power, of relative order $\as^{2/3}$, which come about
because the cutoff introduces an asymmetry: the BFKL evolution can
only take place at scales larger than the cutoff, where the coupling
is reduced by its running.

That the correction should go as $\as^{2/3}$ can be seen as follows.
Recall that in the usual fixed-coupling saddle-point approximation the
gluon Green function between transverse scales $k$ and $k_0$ at
rapidity $Y$ goes as
\begin{equation}
  \label{eq:G}
  G(Y;k,k_0)
  \sim \frac{e^{\omega Y - (\ln^2k^2/k_0^2)/(2\asb \chi''Y) }}{\sqrt{2\pi\asb
    \chi'' Y}}\,,
\end{equation}
where $\omega = \asb \chi(1/2)$, $\asb = \as \NC/\pi$, and $\chi$, the
BFKL characteristic function, and $\chi''$, its second derivative, are
assumed to be evaluated at $\gamma=1/2$, unless otherwise stated. By
examining $\partial_Y \ln G(Y,k,k) = \omega - 1/2Y$, one sees that the
effective BFKL power receives a correction, $\delta \omega$, of order
$1/Y$; the corresponding width in $\ln k^2$ of the solution is of
order $ \sqrt{\asb \chi'' Y}$.  Equivalently if, for `external'
reasons, the width in $\ln k^2$ of the solution is limited to be
$\Delta t$, then the BFKL evolution power will be suppressed by an
amount $\delta \omega \sim (\asb \chi'')/\Delta t^2$ (\cnf
\cite{Collins:1991nk,McDermott:1995jq} for more precise calculations).
When the evolution takes place with a running coupling, the width is
naturally limited by the cutoff in the infrared and by the low value
of the coupling in the ultraviolet. The actual width of the solution
is such that the two sources of suppression, \ie the finite width of
the solution and the running of the coupling, are of similar
importance: $(\asb \chi'')/\Delta t^2 \sim \omega b \asb^2\chi \Delta
t$, where $b = (12\NC-2\nf)/12\pi$. This
gives $\Delta t^3 \sim \chi''/(b\asb \chi)$, or equivalently $\delta
\omega \sim \asb^{5/3} (b\chi)^{2/3} {\chi''}^{1/3}$, \ie a correction
of relative order $\as^{2/3}$. This simple argument actually
reproduces the whole of the suppression's leading functional
dependence on $\as$, $\chi$, $\chi''$ and $b$.

\begin{figure}[htbp]
  \centering
  \includegraphics[width=4.2in]{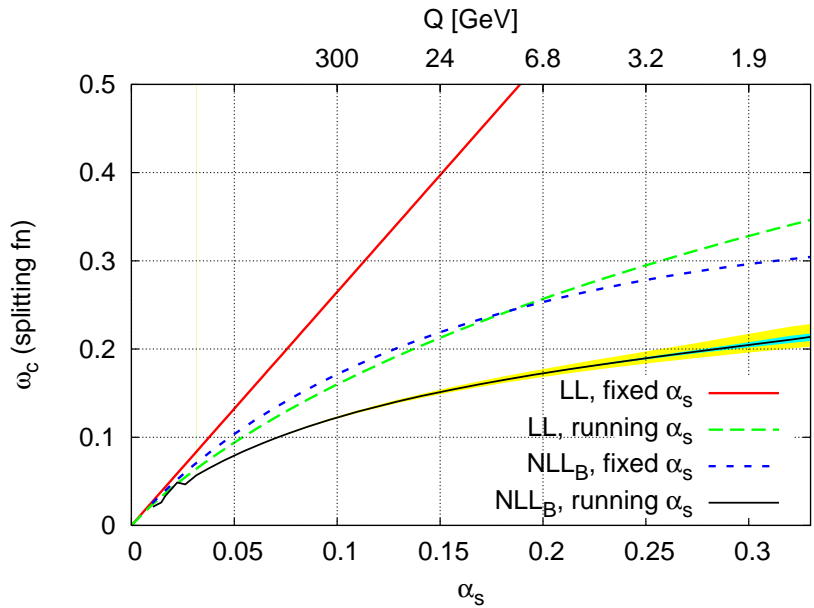}
  \caption{The small-$x$ power growth $\omega_c$ for the $P_{gg}$
    splitting function in various approximations: \LLx and \NLLB
    (\NLLx with additional enhanced higher-order corrections), each
    with fixed and running coupling \cite{CCSSkernel}.}
  \label{fig:omegac}
\end{figure}

Supplementing the above result with the relevant extra numerical
coefficients, and additionally the NLL corrections, the small-$x$
power growth, $\omega_c$, of the $P_{gg}$ splitting function at scale
$Q^2$ becomes
\begin{equation}
  \label{eq:omegacexp}
  \omega_c \simeq  4\ln2 \,\asb(Q^2) \cdot \left(1 - 4.0\asb^{2/3} -
    6.5 \asb  + 
  \order{\asb^{4/3}}\right)
\end{equation}
One sees that, numerically, the running coupling and \NLLx
contributions are both large, negative, and of the same order of
magnitude. In order to make a phenomenological prediction it is
necessary to take into account the running of the coupling at all
orders and to supplement the \NLLx corrections with the yet
higher-order collinear-enhanced terms (which we refer to as \NLLB).
The results for the power are shown in figure~\ref{fig:omegac}. One
sees that despite their different parametric dependence on $\as$, in
practise if one takes individually either the running coupling or the
\NLLB contributions, they lead to almost identical suppressions.
Interestingly though, when taking both running and \NLLB
contributions, there is only limited extra suppression compared to
either one individually.\footnote{\label{ForteFootnote}%
  An alternative way of viewing the results, \cite{FortePersonal}, is to
  consider not the absolute change in $\omega_c$, but rather the
  fractional change as one includes various higher-order contributions
  --- one then notices (with, say, $\as=0.2$) that the inclusion of a
  first higher-order contribution leads to about a $50\%$ reduction in
  $\omega_c$, while the second has almost as large an effect, being a
  further $35\%$ reduction.}

In the context of studies of the evolution of the saturation scale
with both higher-order corrections and running coupling
\cite{Triantafyllopoulos:2002nz}, a similar phenomenon has been
observed, though there, the sequence of results that was shown
corresponded to fixed-coupling \LLx, running coupling \LLx and running
coupling \NLLx\ --- if one considers just this combination of results
then it is tempting (as was done in
\cite{Triantafyllopoulos:2002nz,MuellerTheseProc}) to make the
statement that running coupling effects are dominant and the \NLLx
corrections are rather small. However, recently, the fixed-coupling
(approximate collinearly improved) \NLLx results were presented
\cite{KMRS} for the evolution of the saturation scale, and together
with \cite{Triantafyllopoulos:2002nz}, those results suggest that the
picture is actually very similar to the splitting-function case: each
of \NLLx and running coupling contributions are individually large and
negative, but combining them leads to only a small amount of further
suppression.

Actually, such a result is quite natural: while at the lowest orders,
\eg Eq.(\ref{eq:omegacexp}), different sources of higher-order effects
combine linearly, at higher orders there are strong non-linear
effects.  In the case of running coupling and \NLLx effects there are
actually three physical mechanisms at play: (a) since the cutoff
causes the solution of the BFKL equation to be dominated by higher
scales, where $\as$ is smaller due to its running, \NLLx effects are
reduced; (b) \NLLx effects themselves reduce the dependence of
$\omega_c$ on $\as$, (suppressing it more at large $\as$ than at small
$\as$), slowing the running of $\omega_c$ with transverse scale, as if
there were a reduced `effective' $\beta$-function, and this leads to a
smaller running coupling correction; (c) the \NLLx corrections cause a
very strong suppression of the diffusion coefficient, $\chi''$, which
means that limiting the width of diffusion, as happens due to the
running of the coupling, has a smaller effect on the asymptotic
power.

\begin{figure}[htbp]
  \centering
  \includegraphics[width=4.2in]{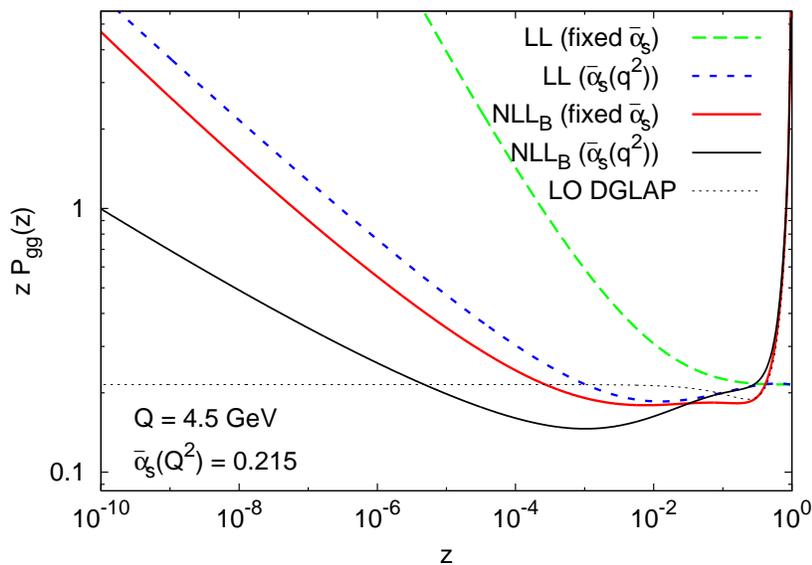}
  \caption{The $P_{gg}$ splitting function with fixed and running
    coupling for the \LLx and \NLLB cases \cite{CCSSkernel}, compared
    to LO DGLAP.}
  \label{fig:Pgg}
\end{figure}

The discussion so far has concentrated just on the power-growth of
splitting functions and saturation scales.  In the case of the
splitting function, with the aid of recent technical developments, it
has become possible to study the whole $x$-dependence of the splitting
function, even at preasymptotic values of $x$
\cite{CCSSkernel,ABF2004}.  Again, one can examine what happens when
switching on, separately, running coupling and \NLLx effects, as shown
in figure~\ref{fig:Pgg}. As was the case when studying just the
asymptotic power, $\omega_c$, one sees that, individually, running
coupling and \NLLx (or rather \NLLB) effects are of similar magnitude.
What should be noted here though, is that when considering the size of
the (phenomenologically relevant) preasymptotic region of $x$ without
growth, there is a rather large additional effect from the
combination of running-coupling and \NLLB contributions --- for
example the point at which the resummed splitting function starts to
become larger than the LO DGLAP splitting function is $x \sim 10^{-1}$
for fixed-coupling \LLx, $x \sim 10^{-3}$ for running-coupling \LLx or
fixed-coupling \NLLB, and $x \sim 10^{-5}$ for running-coupling \NLLB.

Figure~\ref{fig:Pgg} conveys what is perhaps one of the main general
lessons to be retained from studies of resummed splitting functions:
asymptotic properties of small-$x$ resummation have little relevance
at today's energies. This statement holds in two senses: the behaviour
of the splitting function at moderately small values of $x$ is
definitely not power-like; and general properties that one may deduce
from studies of the asymptotic region (\eg that combining \NLLB and
running-coupling effects provides only a modest extra suppression
relative to each one individually) do not hold in the preasymptotic
region. In the case of the splitting functions, the specificity of the
preasymptotic region can be traced to the appearance of new
hierarchies in the perturbative structure, finite towers of terms
$\as^p (\as \ln^2 1/x)^n$, discussed in
\cite{CCSSdip,CiafaloniTheseProc}, leading to the characteristic dip
structure at $\ln 1/x \sim 1/\sqrt{\as} + \order{1}$.
%However, the
%fact that preasymptotic effects are large also for the gluon Green
%function \cite{CCSSkernel,Andersen-SabioVera} suggests that it is a
%quite general feature. \textbf{[not in logic of cutoff effects...]}

In the case of the BK equation, a full study of preasymptotic effects
including higher orders has yet to be carried out.  It would
presumably require that one know the structure of the \NLLx terms not
only for the linear part of the evolution, but also for the non-linear
term.\footnote{Whereas the universality features demonstrated in
  \cite{MunierPesch} ensure that the asymptotic properties of the
  solutions \cite{Triantafyllopoulos:2002nz,KMRS} are independent of
  the details of the non-linear term.} %
Currently however, the higher-order corrections to the non-linear term
are not known. In the meantime it would nevertheless be of interest to
have even just a full study of the $x$ and $Q^2$ structure of the
BK-equation in which only the linear term was supplemented by
higher-order corrections. It is to be noted though that some general
information on the impact of higher-order corrections on
preasymptotics in the BK-equation can already be obtained from studies
\cite{KutakEtAl,Chachamis:2004ab,Gotsman:2004xb} which solve the
BK-equation in $x$, $Q^2$ space with additional terms that partially
mimic the linear \NLLx corrections.

%----------------------------------------------------------------------
\section{Phenomenological impact of resummed splitting functions}

We have seen, Fig.\ref{fig:Pgg}, that preasymptotic effects are large
in the resummation of the gluon-gluon splitting function, so much so
that the BFKL growth only sets in for $z \sim 10^{-5}$. This suggests
that resummation may have only a modest impact on DGLAP fits. To
determine robustly whether or not this is the case would however
require that one carry out a complete DGLAP fit, with not only the
$P_{gg}$ splitting function, but also the whole matrix of splitting
functions and the coefficient functions, preferably in the $\MSbar$
scheme, so as to aid comparison with existing fixed-order DGLAP fits,
\eg \cite{MRSTNLO,MRSTNNLO,CTEQNLO}. This represents a major programme of work,
some aspects of which are currently being investigated.

\begin{figure}[htbp]
  \centering
  \includegraphics[width=4.2in]{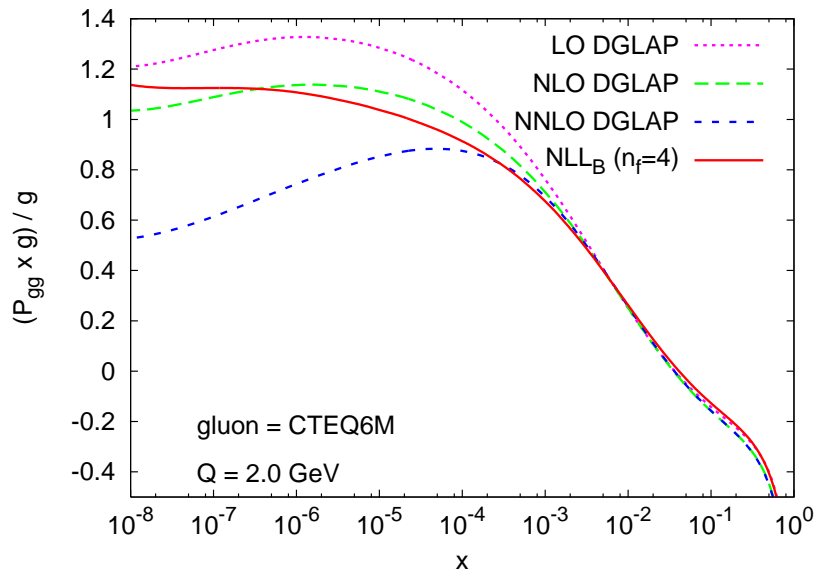}
  \caption{Convolution of CTEQ6M \cite{CTEQNLO} gluon with various
    splitting functions, normalised to the gluon itself,
    $(P_{gg}\otimes g)/g$.}
  \label{fig:convolution}
\end{figure}

Nevertheless, some degree of insight into the possible
phenomenological impact can be obtained simply by taking a fixed gluon
distribution (here CTEQ6M \cite{CTEQNLO}, which has the advantage of
being smooth at small $x$) and examining the convolution
$P_{gg}\otimes g(x,Q^2)$, shown in Fig.\ref{fig:convolution}
normalised to $g(x,Q^2)$. As well as the convolution with the resummed
(\NLLB) splitting function, the plot shows the convolution with the
fixed-order splitting function up to NNLO \cite{NNLO}. The comparison
is to be used only for illustrative purposes since the fixed-order
splitting functions are in the $\MSbar$ scheme (though actually, at
small $x$, the scheme is usually important only starting from
N$^3$LO), while the resummed splitting function is in the $Q_0$
scheme \cite{Q0}.  Furthermore at large $x$ the \NLLB resummation has been
matched only to the LO DGLAP splitting function.

In Fig.~\ref{fig:convolution}, because the gluon distribution itself
rises at small-$x$, a feature of the splitting function at some given
$x$ value manifests itself in the convolution at somewhat smaller $x$.
Thus, though the \NLLB splitting function drops below the LO splitting
function for $x\sim 10^{-1}$ (\cnf Fig.~\ref{fig:Pgg}, though the
$Q^2$ value there is different), this crossover in the convolution
takes place at $x\sim 10^{-2}$. For the crossover in the opposite
direction the effect is much stronger, the \NLLB splitting function
overtaking the LO splitting function at $x\sim 5\cdot10^{-5}$, whereas
in the convolution this occurs below $10^{-8}$.

Looking at the comparison with higher orders, one notices that at
small $x$, the resummed convolution coincides quite closely with the
NNLO convolution --- this is perhaps not unsurprising, since down to
$x\sim 10^{-3}$ there is a good deal of similarity between the NNLO
and resummed splitting functions \cite{CCSSdip,CiafaloniTheseProc}.
Only for $x\lesssim 10^{-4}$ does one start to see a difference
between the NNLO and \NLLB convolutions and, over the remaining
phenomenologically accessible region, the \NLLB convolution is
intermediate between the NLO and NNLO results. If one is courageous
(\ie one believes that the main characteristics will remain the same
after scheme changes and inclusion of the full matrix of splitting
functions and the coefficient functions), one may take this to suggest
that current NNLO fits \cite{MRSTNNLO} should be adequate down to
$x\sim 10^{-4}$ and that only beyond does the fixed-order truncation
truly start to break down.\footnote{If one takes higher-order fixed
  order calculations, they may break down earlier, because large LL
  $(\as^n\ln^{n-1}x)/x$ terms, absent in NLO and NNLO, appear starting
  N$^3$LO.}

% %----------------------------------------------------------------------
% \section{Conclusions}
% \label{sec:outlook}
% 

\smallskip

\textbf{Acknowledgements.} The splitting-function results described
here have been obtained in collaboration with M.~Ciafaloni,
D.~Colferai and A.~Sta\'sto.  I am grateful to the Ettore Majorana
Centre for the kind invitation and financial support to attend this
very enjoyable workshop.

\end{document}